\documentclass[fleqn,usenatbib]{mnras}

\usepackage{newtxtext,newtxmath}

\usepackage[T1]{fontenc}

\DeclareRobustCommand{\VAN}[3]{#2}
\let\VANthebibliography\thebibliography
\def\thebibliography{\DeclareRobustCommand{\VAN}[3]{##3}\VANthebibliography}

\usepackage{graphicx}	
\usepackage{amsmath}	
\usepackage{amssymb}	

\title[FRBs at 408 MHz with the Northern Cross]{The Northern Cross Fast Radio Burst project. I. Overview and pilot observations at 408 MHz}

\author[N.~T.~Locatelli et al.]
{
\parbox{\textwidth}{
Nicola T. Locatelli$^{1,2}$, 
Gianni~Bernardi$^{2,3,4}$\thanks{gianni.bernardi@inaf.it},  
Germano~Bianchi$^{2}$, 
Riccardo~Chiello$^{5}$, 
Alessio~Magro$^{6}$,
Giovanni~Naldi$^{2}$, 
Maura~Pilia$^{7}$, 
Giuseppe~Pupillo$^{2}$, 
Alessandro~Ridolfi$^{7, 8}$, 
Giancarlo~Setti$^{1,2}$, 
Franco~Vazza$^{1,2}$
}
\vspace{0.4cm} \\
\parbox{\textwidth}{
$^{1}$Dipartimento di Fisica e Astronomia, Universit\'{a} di Bologna, Via Gobetti 93/2, 40129 Bologna, Italy\\
$^{2}$INAF-Istituto di Radio Astronomia, via Gobetti 101, 40129 Bologna, Italy\\
$^{3}$Department of Physics and Electronics, Rhodes University, PO Box 94, Grahamstown, 6140, South Africa\\
$^{4}$South African Radio Astronomy Observatory, Black River Park, 2 Fir Street, Observatory, Cape Town, 7925, South Africa\\
$^{5}$University of Oxford, Denys Wilkinson Building, Oxford, OX1 3RH, United Kingdom\\
$^{6}$Institute of Space Sciences and Astronomy (ISSA), University of Malta, Msida MSD 2080, Malta\\
$^{7}$INAF-Osservatorio Astronomico di Cagliari, via della Scienza 5, I-09047 Selargius (Cagliari), Italy\\
$^{8}$Max-Planck-Institut fuer Radioastronomie, Auf dem Huegel 69, D-53121 Bonn, Germany
}
}

\date{Accepted 2020 March 17. Received 2020 March 06; in original form 2020 January 13}

\pubyear{2020}

\begin{document}
\label{firstpage}
\pagerange{\pageref{firstpage}--\pageref{lastpage}}
\maketitle

\begin{abstract}
Fast radio bursts remain one of the most enigmatic astrophysical sources. Observations have significantly progressed over the last few years, thanks to the capabilities of new radio telescopes and the refurbishment of existing ones. Here we describe the upgrade of the Northern Cross radio telescope, operating in the 400-416~MHz frequency band, with the ultimate goal of turning the array into a dedicated instrument to survey the sky for fast radio bursts. We present test observations of the pulsar B0329+54 to characterize the system performance and forecast detectability. Observations with the system currently in place are still limited by modest sky coverage ($\sim 9.4$~deg$^2$) and biased by smearing of high dispersion measure events within each frequency channels. In its final, upgraded configuration, however, the telescope will be able to carry out unbiased fast radio burst surveys over a $\sim 350$~deg$^2$ instantaneous field of view up to $z \sim 5$, with a (nearly constant) $\sim 760 \, (\tau/{\rm ms})^{-0.5}$~mJy rms sensitivity. 
\end{abstract}




\section{Introduction}

Fast Radio Bursts (FRBs) are extremely bright (1-100~Jy), impulsive (0.1-10~ms) transient events dispersed by their propagation through an ionized plasma. Their excess of dispersion measure with respect to the Galactic contribution is  nowadays accepted as a convincing evidence
of their extragalactic origin, but, beyond this, little is still known about their nature and physics \citep[for a review on the topic, see][]{petroff19, cordes19}. Almost one hundred FRBs have been observed to date and only a handful of them appear to repeat \citep{spitler16, CHIME2019b, 2019arXiv190803507T, 2019arXiv190810026K}. A few FRBs have been localized, confirming their extragalactic origin, and their host environments have been found fairly different \citep{2017Natur.541...58C,2018Natur.553..182M,2019Natur.572..352R}. 
This scenario seems to indicate that FRBs may not be a single class of events, and significant effort is nowadays undertaken to localize more bursts \citep{bailes17, bannister19,2019MNRAS.489..919K}.

Beyond localization, the detection of a larger number of FRBs is crucial to discriminate among possible different populations \citep{caleb16, niino18,macquart18II, 2018NatAs...2..865K, james19, locatelli19}, their emission mechanism \citep{lyutikov17, ghisellini17, ghisellini18} and their astrophysical environment (see \citealt{platts18} for an updated review). 
Moreover, a larger statistical sample is necessary in order to use FRBs as effective cosmological probes \citep{mcquinn14, macquart18Nat, akahori16, vazza18FRB, hackstein19, 2019ApJ...872...88R}. 

Initially, FRBs were detected at GHz frequencies \citep{lorimer07, 2013Sci...341...53T, 2014ApJ...790..101S, 2014ApJ...792...19B, 2015MNRAS.447..246P, bhandari18, 2018ApJ...869..181P, shannon18}, but recent observations in the $400-800$~MHz range have enormously increased the FRB statistics \citep[e.g.,][]{caleb16, CHIME18, CHIME2019a} and placed increasingly better upper limits on their event rate \citep{sokolowski18, sanidas19, terVeen19}, showing the advantage of large field of view (FoV) observations.

In this paper we describe the ongoing effort to turn the Northern Cross (NC) radio telescope into a dedicated FRB survey machine observing at 408~MHz. 
We describe the current status of the instrumentation and related observations, and the forecast for upcoming surveys. 
Due to the large FoV of the NC, we expect a detection rate orders of magnitude higher than surveys carried out at GHz frequencies, in particular for distant ($z>2$) events. From our initial estimates we expect to achieve performances comparable to the CHIME/FRB experiment.  

The paper is organized as follows: in \S~\ref{sec:instrument} we describe the current instrument status and recent upgrade, in \S~\ref{sec:test_obs} we present test observations that characterize the system, 
in \S~\ref{sec:survey_design} we forecast the FRB detection with the NC and we conclude in  \S~\ref{sec:conclusions}.

\section{Instrument description} \label{sec:instrument}
\begin{figure}
	\includegraphics[width=\columnwidth]{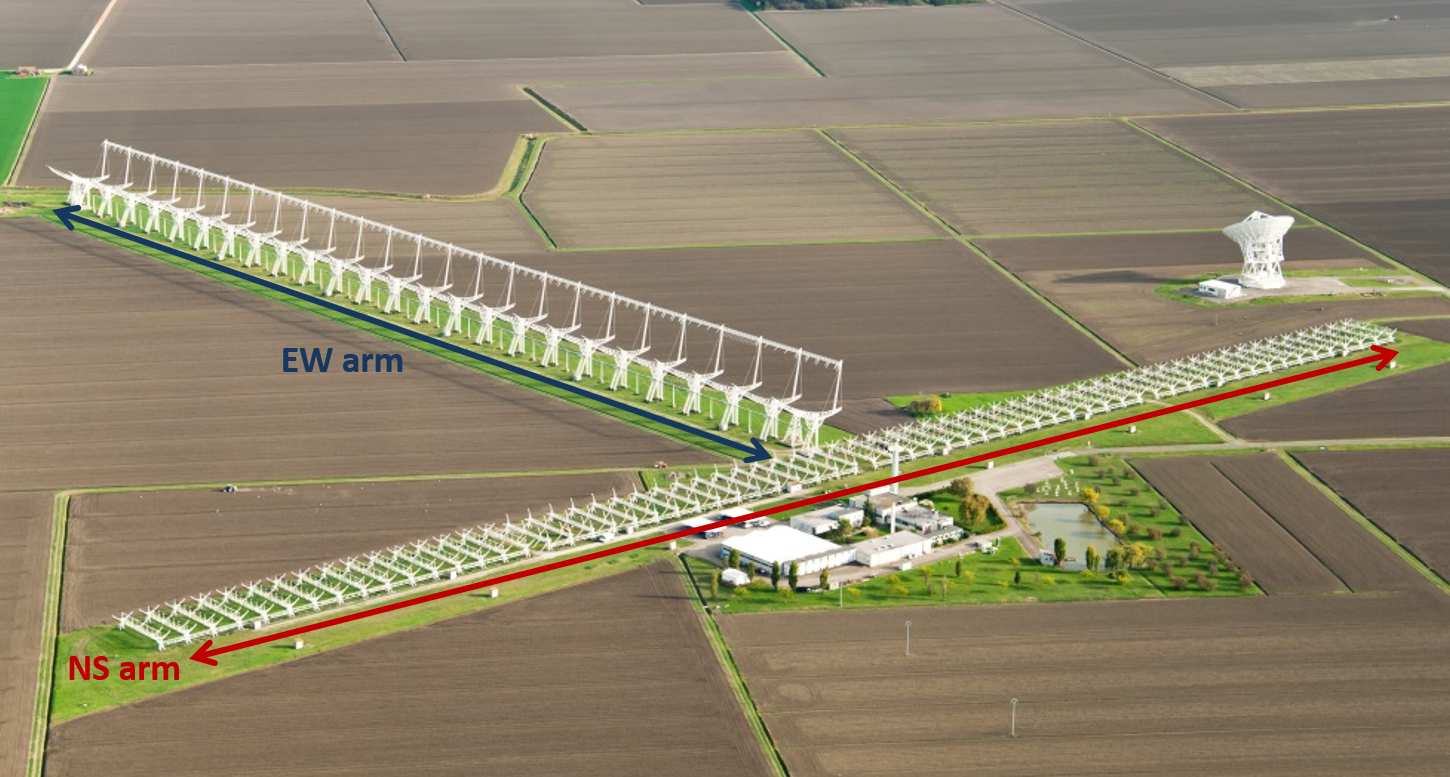}
    \caption{Aerial view of the Medicina Radio Astronomical Station. In the foreground, the Northern Cross with its two orthogonal arms.}
    \label{fig:Med_station}
\end{figure}
The NC is a T-shaped radio interferometer operating at 408~MHz, located at the Medicina Radio Astronomical Station (Bologna, Italy). Its orthogonal arms are aligned along the North-South (NS) and East-West  directions respectively (Figure~\ref{fig:Med_station}). 
Historically, the NC was used to survey the sky, producing several catalogues of extragalactic radio sources \citep[e.g.,][]{colla70, ficarra97, pedani99}.

The NS arm has 64 reflective cylinders, {$7.5 \times 23.5$~m} each, for a total collecting area $A_{\rm NS} = 11280$~m$^2$. However, as the antenna efficiency is $\sim 0.71$ \citep{bolli08}, the effective area is reduced to $A_{\rm NS,eff} \simeq 8000$~m$^2$.
Each cylinder focuses the incoming radiation on 64 dipoles placed on the focal line; cylinders are spaced 10~m apart, leading to a total arm length of 640~m. The East-West arm was not used in this work.

\subsection{Antenna and analogue receiver upgrade}

\begin{figure}
	\includegraphics[width=\columnwidth]{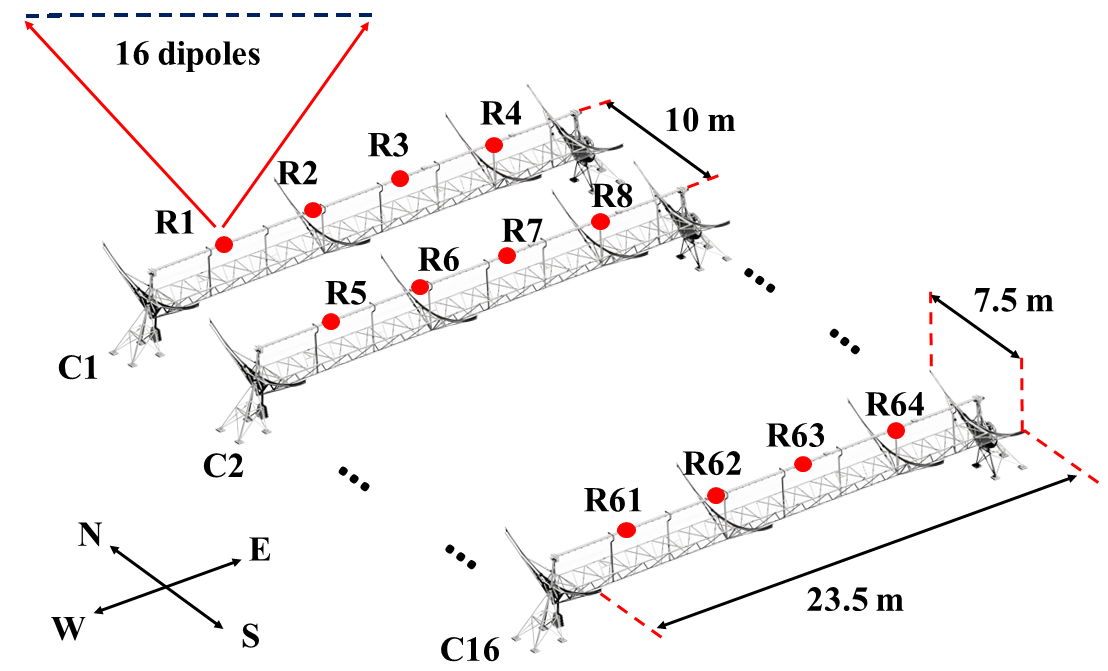}
    \caption{Scheme showing the refurbished section of the NS arm. The red circles indicate the position of the four receivers located on the focal line of each cylinder.}
    \label{fig:NC_Section_Scheme}
\end{figure}

The NS arm is undergoing an upgrade of the antenna and receiving system \citep[see][for details]{SMontebugnoli2009_1}. The focal line of sixteen cylinders has been modified in order to group the signals of sixteen dipoles together, providing four analogue signals per cylinder, i.e. 64 receiving inputs for the refurbished sector (Figure~\ref{fig:NC_Section_Scheme}).
Each receiving input (hereafter only receiver) is connected to a front-end box, mounted on top of the focal line, hosting a low noise amplifier and an optical fibre transmitter  \citep{FPerini2009_1}. The amplified Radio Frequency (RF) signals are sent to the station building through analogue optical fibre links \citep{FPerini2009_2}. The RF receiver includes the optical-electrical conversion, filtering, amplification, conditioning and single down-conversion to the Intermediate Frequency (IF) of 30 MHz \citep{FPerini2009_3}. The output power can be digitally attenuated up to 31.5~dB in steps of 0.5~dB.
A splitter chain architecture is used to distribute the local oscillator (378~MHz), clock and synchronization signals to the IF circuitry.

\subsection{Digital backend}

\begin{figure}
	\includegraphics[width=\columnwidth]{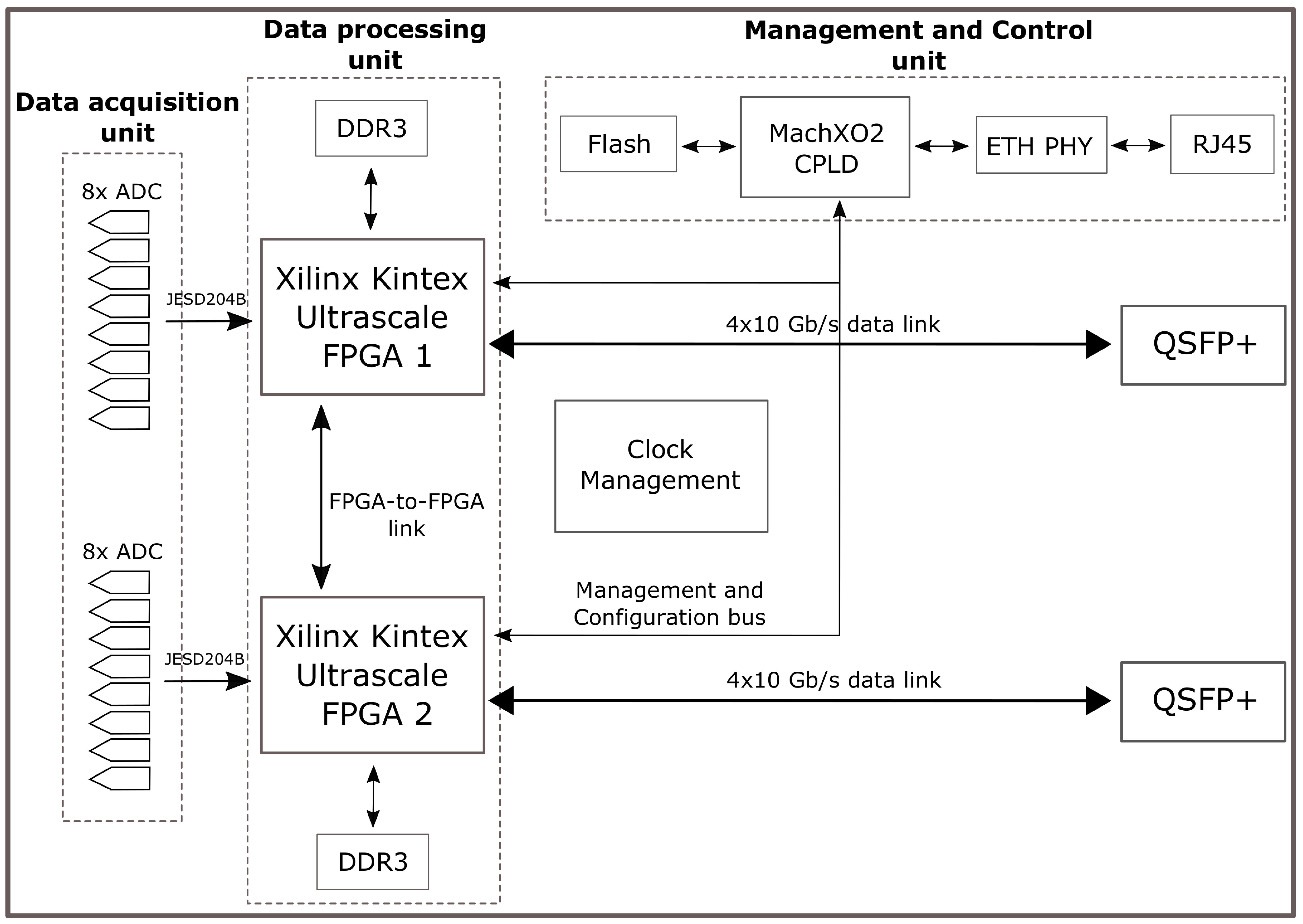}
    \caption{Block diagram showing the main functional parts of the ADU board.}
    \label{fig:adu_scheme}
\end{figure}

The digital acquisition and signal processing hardware is based on the Analog Digital Unit board \citep[ADU,][Figure~\ref{fig:adu_scheme}]{Naldi2017_tpm_hw}, a digital platform developed for the Low Frequency Aperture Array (LFAA) component of the Square Kilometre Array.
The ADU consists of sixteen dual-input Analog to Digital Converters (ADCs) and two Field Programmable Gate Array (FPGA) devices, capable of digitizing and processing the broadband (up to 400 MHz bandwidth) RF streams from 32 single polarization (or 16 dual polarization) antennas at an 800 MHz sampling rate.
The 32 RF analogue inputs are digitised by sixteen 14-bits dual-input ADCs AD9680 that send the eight most significant bits to the XCKU040 FPGAs. These samples are time stamped using a pulse-per-second signal with a reference high precision clock, and the ADU is synchronised to Coordinated Universal Time via the control interface.

The firmware design is highly modular, with a board specific I/O ring containing the interfaces to the physical peripherals and the control structure, and a core containing the signal processing chain \citep{Comoretto2017_tpm_fw}. The board is controlled using an AXI4lite bridge to the 1~Gb Ethernet port, with each element seen as a memory mapped portion of the board address space. A map of this space is generated automatically at compile time and used by the control software to address each element by name (\citealt{aavsaccess}).

Signal processing is performed on the FPGAs, with the resulting output data transferred to a processing server through a 40~Gb Ethernet connection. 
\begin{table}
	\centering
	\caption{Characteristics of the current acquisition system.}
	\label{table:dbe_properties}
	\begin{tabular}{l c}
        \hline\hline 
        N. of frequency channels & 1024\\
        Channel width & 781.25 kHz\\
        Time resolution & $1.08~\mu$s\\
        \hline
        \multicolumn{2}{c}{Multibeam beamformer} \\
        \hline
        N. bits & 16 complex\\
        N. channels & 384 \\
        N. beams & 4\\
        Max. time resolution & $69.12~\mu$s\\
        Max. throughput & 355.56 Mb/s\\
        \hline
        \multicolumn{2}{c}{Single beam beamformer} \\
        \hline
        N. bits & 16 complex\\
        N. channels & 21\\
        Throughput & 311.11 Mb/s\\
        \hline
    \end{tabular}
\end{table}
The signal processing chain includes:
\begin{itemize}
    \item Correction for cable mismatch: Relative delays due to cable mismatch can be compensated for by applying a time-domain shift to each of the 24 IF inputs;
    \item Channelization: Each of the 24 IF inputs are channelised into 512, 781~kHz-wide channels by an oversampled polyphase filterbank;
    \item Correction of instrumental and geometric delays: Combined calibration and pointing coefficients are provided to the FPGAs.  A calibration coefficient per antenna and per channel is required to correct for the receiver amplitude and phase response. Separate pointing coefficients are required for each generated beam, such that each beam can be pointed independently; 
    \item Frequency domain beamforming: The signal processing firmware can generate four simultaneous beams with a minimum integration time of $\sim 70$~$\mu$s and one beam at 1.08~$\mu$s time integration. During this stage, the coefficients provided in the previous step are applied to each channelised data stream, thus simultaneously calibrating and pointing each beam.
\end{itemize}

The main features of the acquisition system are summarized in Table~\ref{table:dbe_properties}.
The ADU is managed using the monitoring and control software prototyped for the LFAA and uses the UniBoard Control Protocol for communication between the board and a compute server \citep{aavsaccess}. The management software runs on the server and can read from and write to the FPGA memory address and other devices on the board. Management operations include: programming of the FPGAs; on-board device control; FPGA and ADC synchronization; network setup; configuration of the signal processing chain; download of calibration and pointing coefficients for each beam; and instructions to broadcast control and data products. 
All processing on the ADU  (i.e., from digitization to the transmission of beamformed data) is performed in real-time.

The generated beams are transmitted from the ADU to the server over a 40~Gb link using a custom SPEAD protocol. A subset of the raw channelized data (the output of the firmware channelizer) can also be directly broadcast and used to generate calibration coefficients. Additional control data streams include transmission of raw antenna voltages and integrated spectra per antenna, both used to monitor the system performance. 

Data streams are received at the compute server using the data acquisition system developed for the LFAA \citep{aavsdaq}. The data acquisition system can process different simultaneous data streams concurrently. A "packet consumer" is associated with each stream type, such that two specialised consumers are required: one to receive the fully sampled beam; and another to receive the raw channelised data. The channelized data streams are stored to disk using a simple binary format and are then correlated to generate calibration coefficients (see \S~\ref{sec:test_obs}). The frequency channels of interest from the fully sampled beam are saved to disk using a modified version of the SigProc Filterbank  file format \citep{sigproc}, where the complex voltages, rather than the power, are stored. This modification reduces the processing requirements (i.e. eliminates per-sample processing), resulting in the system being capable of writing data to disk in real time, and allows for custom offline software to convert the file to different file types such that no signal information is lost. For the tests described in this paper, custom filterbank files are converted to filterbank compatible files.

The system (front end and back end) described above is already a major upgrade over the pulsar back end used in the late '90s for pulsar searches and timing \citep{damico96}, however, we have already started to further optimize the system for FRB observations. In particularly, upcoming upgrades will include  digitization at 700 MS/s, thus sampling the RF analogue band in the second Nyquist zone. The new design will implement a Digital Down Converter (DDC) that down-converts the signal of interest to base-band and filters out the image band that originates from the mixing operation. The sampling rate will be reduced by approximately two orders of magnitude and the channelizer modified accordingly, while maintaining the same oversampled polyphase filterbank structure. 
The beamformer will be re-designed in order to  produce up to twenty independent beams, placed anywhere inside the single element FoV.

We are currently working on developing an online FRB search pipeline that performs the standard steps of de-dispersion, candidate identification and storage for further reprocessing \citep[following a scheme similar to][for example]{CHIME18}, building on the HEIMDALL\footnote{https://sourceforge.net/p/heimdall-astro/wiki/Use/} publicly available code \citep[e.g.,][]{gajjar18}.

\section{Test observations} 
\label{sec:test_obs}

We performed test observations in order to validate the system for FRB studies. As described in \S~\ref{sec:instrument}, the digital beamformer requires that the receiver signals are corrected for the corrupting effects that arise along the RF path. This calibration procedure is done through standard interferometric techniques where the channelized complex voltages $v$ from each receiver pair $(i,j)$ are recorded and cross correlated to form visibilities $V_{ij}$:
\begin{equation}
V_{ij} = \langle v_i(t) \, v_j(t)^* \rangle_{\Delta t},
\label{eq:correlation}
\end{equation}
where $\langle \rangle_{\Delta t}$ indicates the average over the integration time  ${\Delta t}$ and $^*$ is the complex conjugate.
A software correlator is used to evaluate the right hand side of equation~\ref{eq:correlation} by integrating the cross products over $\Delta t = 1.13$~s, that is a trade-off between signal-to-noise ratio (SNR) and fringe smearing.

\begin{figure}
	\includegraphics[width=\columnwidth]{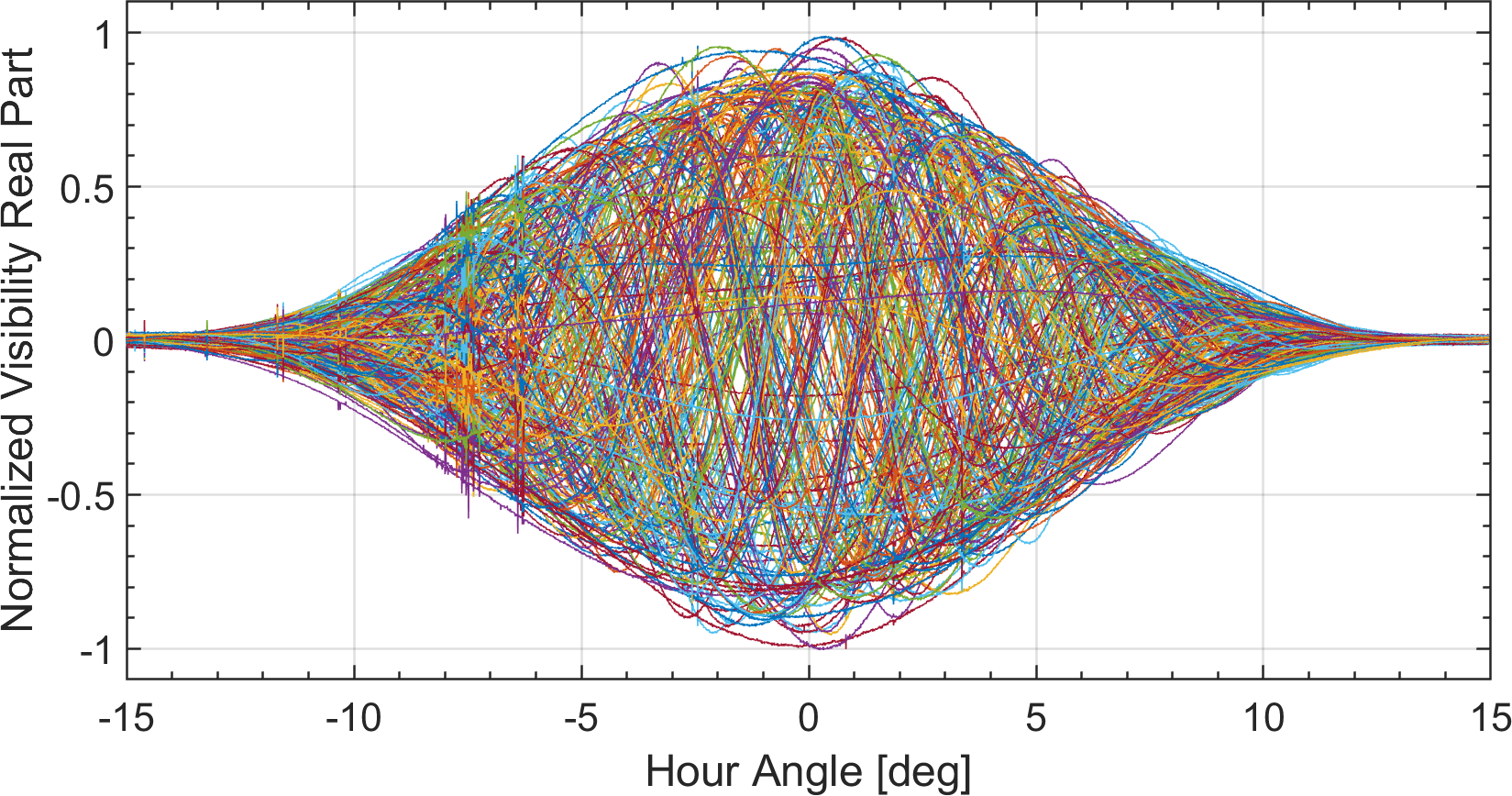}
	\includegraphics[width=\columnwidth]{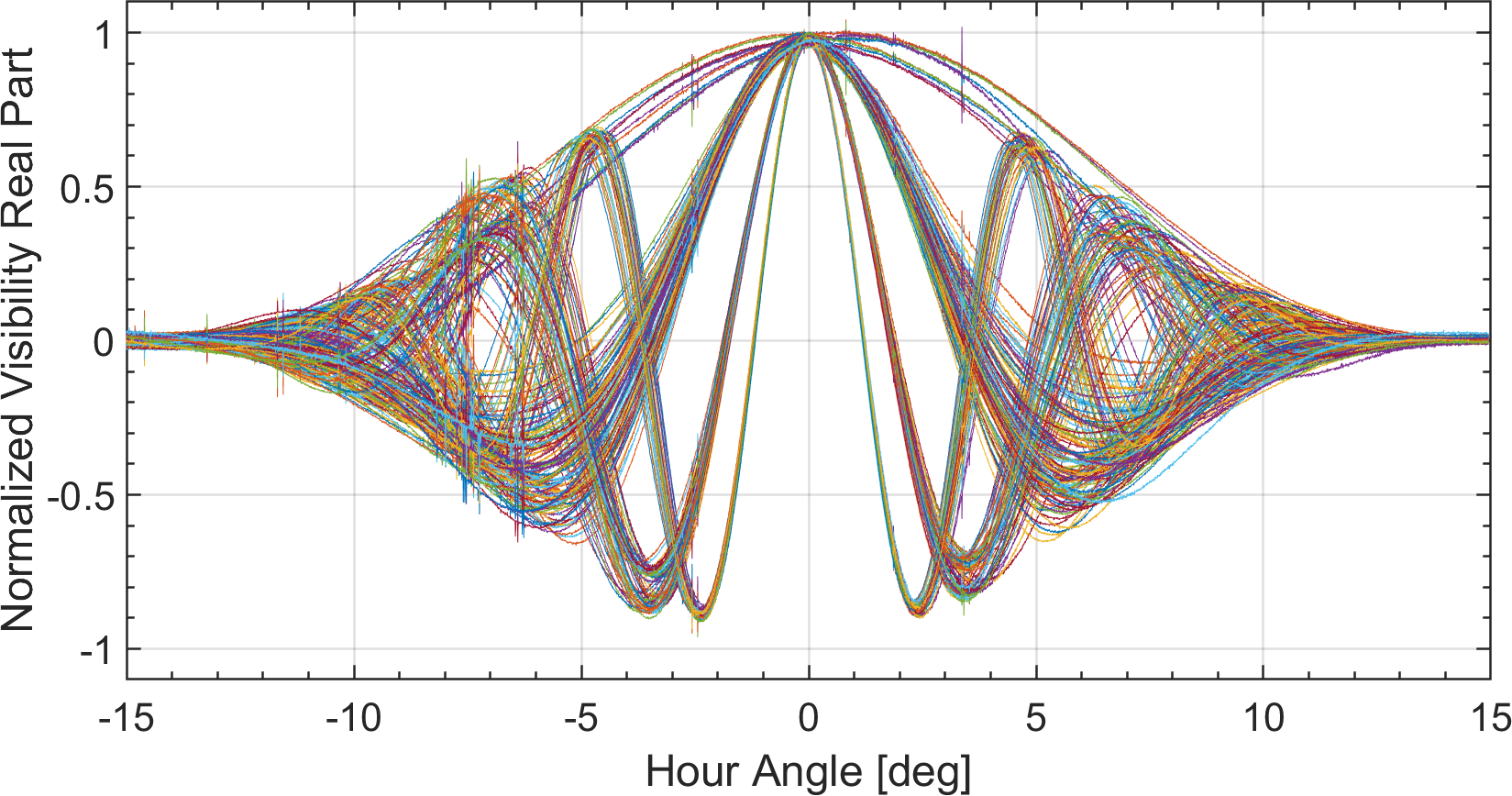}
    \caption{Real part of the complex visibilities corresponding to the transit of  Cas~A before (top panel) and after (bottom panel) calibration. Only the central channel at 407.6875 MHz is shown. Fringes show the main peak at transit ($\omega = 0^\circ$) where they are phased, i.e. where the geometrical delay is compensated. Each colour corresponds to one of the 276 independent baselines. Fringe spacings are proportional to baseline lengths, therefore redundant visibilities appear grouped in subsets that have similar fringe frequencies.}
    \label{fig:CasA_Fringes}
\end{figure}

The instrumental corruptions can be described by complex receiver gains $g$: 
\begin{equation}
V_{ij}^{\rm o}(t,\nu) = g_{i}(t,\nu) \, g_{j}^*(t,\nu) V_{ij}(t,\nu),
\end{equation}
where $V^{\rm o}$ are the observed visibilities, i.e., the visibilities that are corrupted by the instrumental response.
The calibration procedure involves determining the instrumental gains $g$ that can be solved for if the visibilities $V_{ij}$ are known, i.e., through the observation of a calibration source. We observed Cas~A, a standard calibrator for which we assumed a 4467~Jy flux density at 408~MHz \citep{perley17}. Observations were carried out for $\sim 2$~hours in the single beam mode (details are reported in Table~\ref{table:RF_properties}). Six cylinders are formed by a total of 24 receivers, leading to 276 independent baselines, most of which are redundant due to the regular configuration grid (see Figure~\ref{fig:NC_Section_Scheme} for a reference scheme of the array used).
\begin{table}
	\centering
	\caption{Specifications of the NC test observations.}
	\label{table:RF_properties}
    \begin{tabular}{l c}
        \hline                  
        Central observing frequency & 408 MHz\\
        Analogue bandwidth & 16 MHz\\
        Total number of cylinders & 6\\
        Total number of receivers & 24\\
        Longest baseline (NS) & 50 m\\
        Receiver FoV & $\sim38$~
        deg
        $^2$\\
        Receiver FoV FWHM North–South & $5.9^\circ$\\
        Receiver FoV FWHM East–West & $6.4^\circ$\\
        \hline
    \end{tabular}
\end{table}
Visibility data were edited and flagged, and calibration equations solved using two different minimization methods \citep{boonstra03}, obtaining consistent solutions. 
Examples of visibilities compensated for the delay corresponding to the position of Cas~A at the local meridian are shown in Figure~\ref{fig:CasA_Fringes}. 
The bottom panel clearly shows that, after calibration, the real part of the visibilities has maxima aligned in the desired direction, at hour angle $\omega = 0^\circ$.  

We used the derived antenna gains, combined with the geometric delay compensation coefficients, to beamform the six cylinder array towards the pulsar PSR\,B0329+54 \citep{cp68}. PSR\,B0329+54 has a 714~ms period \citep{hobbs04}, a $S_{400} = 1500$\,mJy flux density at 400~MHz \citep{lorimer95} and a dispersion measure, ${\rm DM} = 26.7641$~pc~cm$^{-3}$ \citep{hassall12}. It was observed for $\sim 20$~minute around transit.

We analyzed 20~s-long observations using the standard DSPSR \citep{vanstraten11}, PSRCHIVE \citep{hotan04} and PRESTO \citep{ransom01,ransom02} suites for de-dispersion, folding and Radio Frequency Interference (RFI) excision. The time series was de-dispersed and the single-pulse signal was folded over the integration. Our observations revealed a fairly benign RFI environment. A negligible fraction of the data was visually identified as corrupted and manually flagged. 

Our observing band is partially ($406-410$~MHz) protected and reserved to radio astronomy and partially ($400-406$ and $410-416$~MHz) is assigned to the Italian Minister of Economic Development, therefore we do not expect to have persistent but rather negligible RFI contamination as we found here. There is a regular RFI monitoring programme running at the Medicina station that shows a fairly low RFI occupancy. Interference signal mostly occur from atmospheric balloon probes but are confined to the low part of the band and limited at specific time intervals. Radio link transmissions are also generally short and sporadic. Nevertheless, we intend to implement more automatic RFI flagging strategies, ranging from simple amplitude clipping and standard deviation outliers, to machine learning classification \citep[e.g.,][]{pedregosa12} and use spatially neighbouring beams to discriminate between man-made and sky signal \citep[e.g.,][]{CHIME18,bailes17}.

\begin{figure}
	\includegraphics[width=\columnwidth]{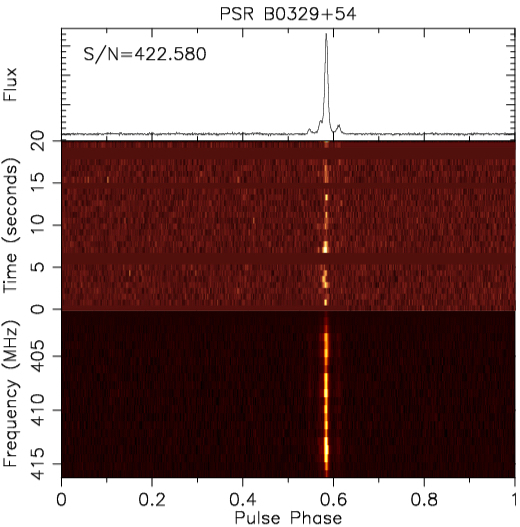}
    \caption{Observed profile of B0329+54. 
    Bottom panel: intensity profile as a function of rotational phase and channel width, integrated over 20~s. We note a slight decrease at the band edges due to the sensitivity loss. 
    Central panel: intensity profile as a function of a single-pulse time ($\sim 700$~ms) over the 16~MHz bandwidth.
    The blanked horizontal bands represent time affected by RFI and, therefore, discarded. We note that no further flagging was needed. 
    Top panel: pulse profile integrated over frequency and 20~s.}
    \label{fig:B0329_20s}
\end{figure}

PSR\,B0329+54 was visible in each 20~s observation, and we used data taken closest to transit to estimate a ${\rm SNR} \sim 422$ (Figure~\ref{fig:B0329_20s}), which, in turn, implies an rms noise $\sigma_6 = \frac{S_{400}}{\rm SNR} \sim 3.6$~mJy - where $\sigma_6$ indicates the sensitivity of the six cylinder array. 

The derived sensitivity can be used to determine the System Equivalent Flux Density (SEFD) of a single receiver, which is the quantity that we ultimately want to characterize. The receiver sensitivity $\sigma$ is given by:
\begin{equation}
\sigma = A \, \sigma_6 
\label{eq:sens_scaling}
\end{equation}
where $A = 24$ is the ratio between the area corresponding to six cylinders and one receiver respectively. The receiver SEFD is then given by the radiometer equation \citep[for a similar approach, see][]{amiri17}:
\begin{equation}
{\rm SEFD} = \sigma \sqrt{N_p \, B \, t} = A \sigma_6 \sqrt{N_p \, B \, t}, 
\label{eq:radiometer_inv}
\end{equation}
where $N_p$ is the number of polarizations measured, $B$ the bandwidth and $t$ the observing time. In our case we have $N_p = 1$, $B = 16$~MHz, $t = 20$~s, obtaining ${\rm SEFD} \sim 1530$~Jy. 

\section{FRB survey design} \label{sec:survey_design}

The system characterization allows us to forecast the FRB detectability with the NC. The telescope can already be used to observe known - i.e. repeating - FRBs, but, given its large FoV, it is best suited to carry out blind surveys to detect new FRBs. 

The NC cylinders can be synchronously steered in declination by a common driveshaft that can be disabled, allowing each cylinder to be moved independently. The elevation range that can be observed without shadowing spans $45^\circ$ from zenith, therefore $0 < \delta < 90^\circ$ is the maximum observable declination range. Recalling that the receiver FoV is $\sim 6^\circ$ wide, 15~pointings are needed to cover the $90^\circ$ declination interval. We therefore envisaged three different modes to observe FRBs with the NC: 
\begin{itemize}

\item [I] {A pilot blind survey: the hardware and software upgrade described in \S~\ref{sec:instrument} has been completed for eight cylinders which can, in turn, be split in two groups of four cylinders, each pointing $6^\circ$ apart. With the current beamformer, each pointing can be tiled with four beams, each $1^\circ \times 1.6^\circ$ wide, placed along the right ascension direction. Such survey will cover $A_{\rm FoV} \sim 9.4$~deg$^2$ instantaneously with a sensitivity $\sigma_I$:
\begin{equation}
\sigma_I = \frac{\rm SEFD}{A_{16} \sqrt{B}} \sim 760 \, (\tau/{\rm ms})^{-0.5} \, {\rm mJy}
\end{equation}
where $A_{16} = 16$, i.e. the number of receivers corresponding to four cylinders and  $\tau$ is the observed time expressed in milliseconds (see also equation~\ref{eq:tau_obs} below);}

\item [II] {A blind survey that covers the widest possible area: once the whole NS arm is upgraded, the layout of the pilot blind survey can be extended to observe the whole declination range that can be accessed instantaneously, i.e. $0 < \delta < 90^\circ$, covered by fifteen pointings spaced $6^\circ$ apart. Each pointing is observed with four cylinders, i.e. leading to the same sensitivity as per the survey I. The current system cannot take full advantage of the increased sky coverage as the four independent beams only cover $\sim 10\%$ of the receiver FoV. For this survey we therefore considered that the improved multibeam and channelization capabilities anticipated in \S~\ref{sec:instrument} are already deployed on all the sixty cylinders.
If twenty independent beams are independently placed within the receiver FoV, the instantaneous sky coverage improves dramatically to $A_{\rm FoV} \sim 350$~deg$^2$. We will use this layout as our best case for FRB observations;}

\item Follow up of known (repeating) FRBs. Known sources can be followed for $\sim 30$~minutes as they transit through the receiver FoV. If sixty cylinders are beamformed together in a $4.5' \times 1.6^\circ$ beam, a $\sigma_{60} \sim 50 \, (\tau/{\rm ms})^{-0.5}$~mJy sensitivity can be achieved.
\end{itemize}
A limitation of the current acquisition system is the relatively coarse frequency resolution that can lead to time smearing of high DM events.
For a transient event of intrinsic duration $t_i$, equal or shorter than the sampling time $\Delta t_{b}$, the observed time $\tau$ is defined as \citep[e.g.,][]{amiri17}:
\begin{equation}
    \tau = \sqrt{\Delta t_{b}^2 + t_{s}^2 + t_i^2}, 
\label{eq:tau_obs}
\end{equation}
where $t_s$ is the scattering time and $t_i$ is the intrinsic time duration of the event. 
If the signal propagates through an ionized plasma, it experiences an additional dispersion delay $t_{\rm DM}$ so that:
\begin{equation}
    \Tilde{\tau} = \sqrt{\Delta t_{b}^2 + t_{s}^2 + t_i^2 + t_{\rm DM}^2}, \label{eq:tau_obs_disp}
\end{equation}
where the dispersion smearing is \citep[e.g.,][]{2014ApJ...792...19B}:
\begin{equation}
    t_{\rm DM} = 8.3 \frac{\rm DM}{\left[ \frac{\rm pc}{{\rm cm}^{-3}} \right]} \frac{\Delta \nu_{\rm ch}}{[\rm MHz]} \left( \frac{\nu}{\rm [GHz]}\right)^{-3} \, \mu {\rm s}, \label{eq:time_smearing}
\end{equation}
where $\Delta \nu_{\rm ch}$ is the channel width and $\nu$ is the observing frequency. 
With the current system, an FRB with a DM~=~647~pc~cm$^{-3}$ (the mean of the known FRB population to date, consistent with 665~pc~cm$^{-3}$ as estimated by \citealt{bhandari18}) would experience an intra-channel dispersion (smearing) $\Tilde{\tau}$:
\begin{equation}
\Tilde{\tau} \simeq t_{\rm DM} \sim 62 \, {\rm ms}, 
\label{eq:condition}
\end{equation}
that becomes 248~ms for the highest DM observed to date, 2596~pc~cm$^{-3}$ \citep{bhandari18}.  
A smaller channel width reduces the intra-channel smearing, normally implying an increase of the sampling time that, however, needs to remain sufficiently small to properly sample the burst duration. We quantified the impact of the  intra-channel smearing for the surveys I and II by estimating the FRB event rate
following \cite{connor19}. Table~\ref{tab:telescopes} summarizes the main specifications of both surveys, where, like we defined above, survey II already employs the finer channelization anticipated in \S~\ref{sec:instrument}.

\begin{figure}
	\includegraphics[width=\columnwidth]{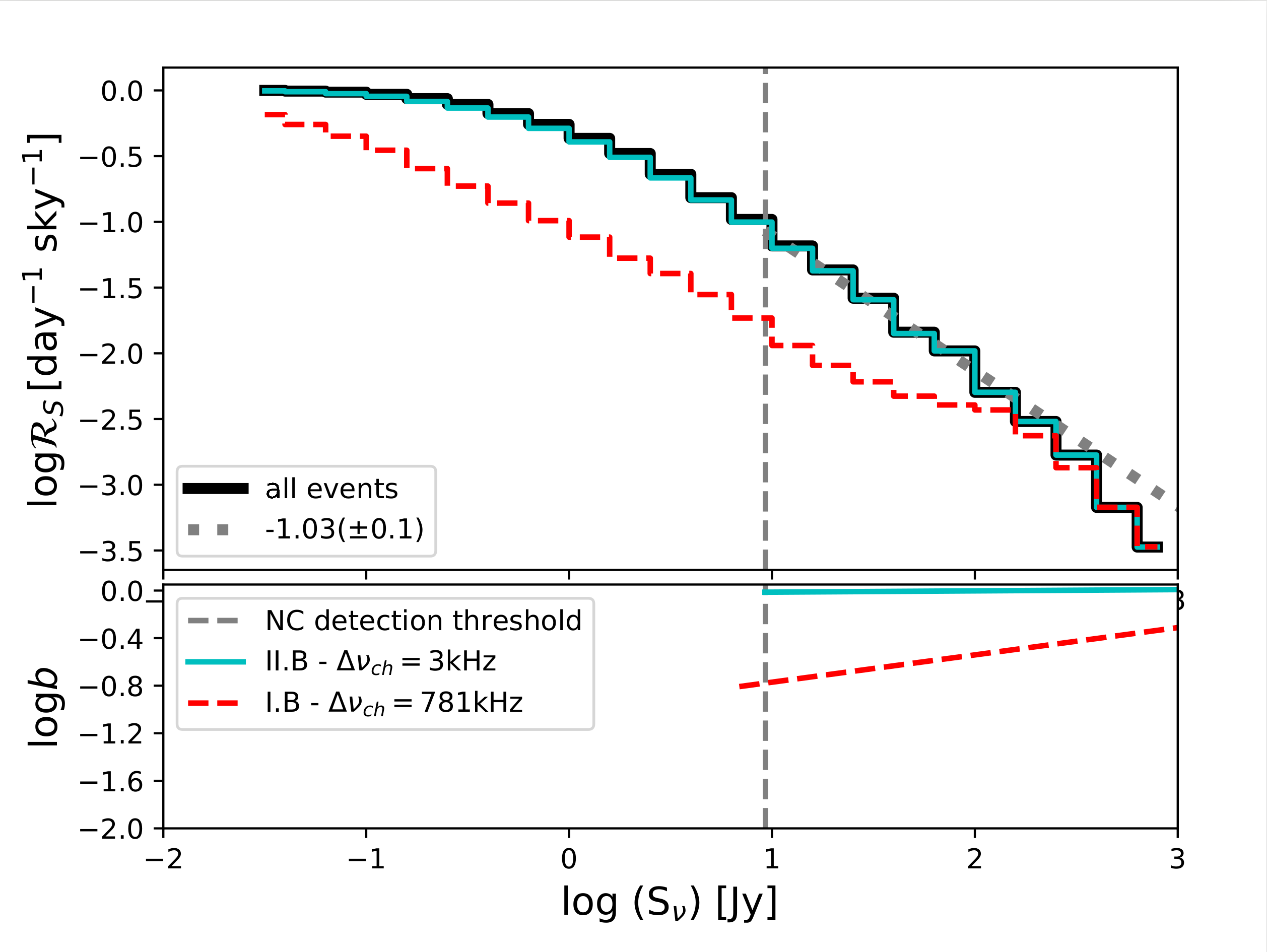}
    \caption{{\it Upper panel}: cumulative event rates $\mathcal{R_S}$ normalized to their relative peaks. 
    The black solid line represents the theoretical prediction for the ideal case with no intra-channel smearing. The curve was fitted by a power law above the detection threshold. The best-fit power law is plotted as a grey dotted line and its slope is reported in legend.
    The dashed red line and the solid cyan line show case I and II from Table~\ref{tab:telescopes} respectively. The cyan and black lines are virtually overlapping. The vertical dashed line represent the $10\sigma_I$ detection threshold - which is the same for both surveys. 
    {\it Lower panel}: bias parameter $b$ as a function of flux density (see text for details).}
    \label{fig:rates_CSFR}
\end{figure}

\begin{figure}
	\includegraphics[width=\columnwidth]{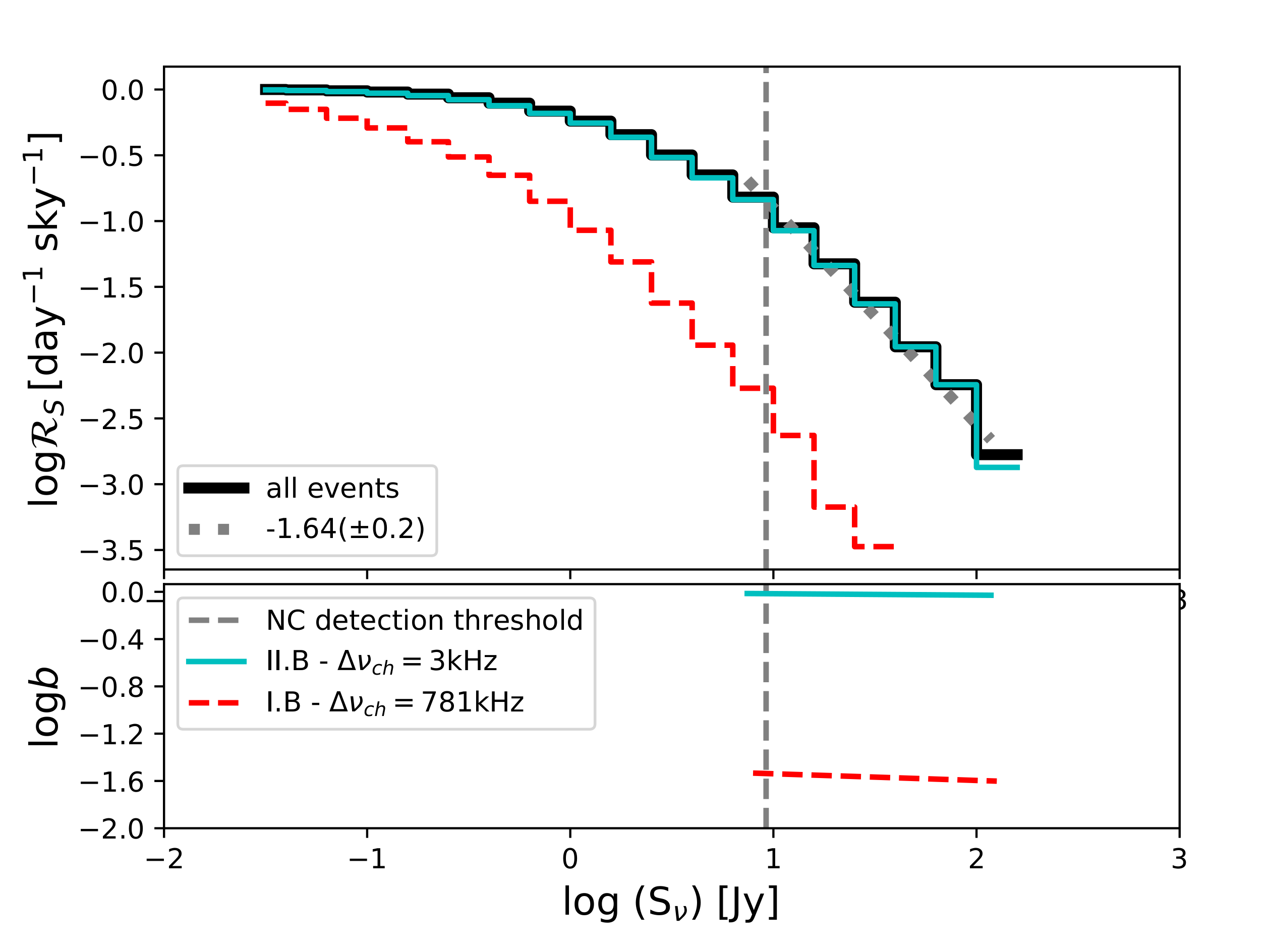}
    \caption{Same as Figure~\ref{fig:rates_CSFR}, for the FRBFR model.}
    \label{fig:rates_FRBFR}
\end{figure}

Event rate estimates require the knowledge of the FRB cosmological distribution, their spectral index, their distribution in duration and their intrinsic luminosity.
In particular, we adopted the following assumptions:
\begin{itemize}
    \item a linear relation between the FRB dispersion measure and its redshift, i.e. DM~$ = 1000 \, \, z$~pc~cm$^{-3}$ \citep{inoue12, dolag15, 2018NatAs...2..865K, 2019MNRAS.487.3672Z}; 
    \item a log-normal distribution for the FRB luminosity function at 1.4~GHz $L_{\rm GHz}$, peaking at $10^{33}$~erg~s$^{-1}$ and full-width at half maximum of 1.5;
    \item a constant spectral index $\beta = 1.5$ for each event\footnote{$S_\nu \propto \nu^{-\beta}$, where $S_\nu$ is the flux density at the frequency $\nu$.}, consistent with the average spectral index of known FRBs  \citep{macquart19}. Although this assumption is likely incorrect, it only affects the rates observed at different frequencies and not the rates observed by the two surveys. 
\end{itemize}
We assumed that the FRB cosmic evolution either follows the cosmic star formation rate \citep[CSFR,][]{madau14}, or a phenomenological formation rate \citep[FRBFR,][]{locatelli19}.
In their work, \cite{locatelli19} model the FRB cosmological evolution following the observed distribution of the events with ${\rm DM} \lesssim 1000$~pc~cm$^{-3}$ \citep{shannon18, macquart18Nat}. In this model, the evolution is faster than the CSFR model and peaks at earlier redshifts.
The cumulative event rate $\mathcal{R_S}$ above a given flux density threshold is shown in Figure~\ref{fig:rates_CSFR} (Figure~\ref{fig:rates_FRBFR}) for the CSFR (FRBFR) model. 
\begin{table}
	\centering
	\caption{NC parameters for the two proposed surveys (see text for details): survey type; sampling time; channel width; instantaneous sky coverage; expected noise level (per millisecond).}
    \label{tab:telescopes}
     \begin{tabular}{l c c c c c c c}
        Survey & $\Delta t_{\rm b}$ & $\Delta\nu_{\rm ch}$ & A$_{\rm FoV}$ & $\sigma$ \\
        type & $\mu$s & kHz & 
        deg
        $^2$ & mJy~$(\tau/{\rm ms})^{-0.5}$\\
        \hline
        I & 70 & 781 & 9.4 & 760 \\
        II & 276 & 3 & 350 & 760 \\
        \hline
    \end{tabular}
\end{table}
We also calculated the bias parameter $b$:
\begin{equation}
    b \equiv \frac{ \mathcal{R}_S (\Delta \nu_{ch})}{\mathcal{R}_S (\Delta \nu_{ch} \xrightarrow{} 0)},
\end{equation}
i.e. the ratio between a given rate and the ideal rate - i.e., the rate unaffected by intra-channel smearing. 
We note that FRBs have a noticeable spectral modulation at low frequencies \citep{CHIME18,CHIME2019a,CHIME2019b}, however, this effect is less prominent for the relatively narrow band of our observations, compared to wider bandwidth instruments.

We expect a significant loss of events due to intra-channel smearing for the survey I, with a magnitude that depends upon the chosen FRB model. In the CSFR case, there is  essentially no event loss at the bright end of the cumulative event rate, whereas the completeness decreases to 17\% at the detection threshold. For the FRBFR case, the loss is already significant for bright events. The reason for this difference is due to the fact that low-redshift events have a higher DM in the FRBFR model than the CSFR one, leading to a higher intra-channel smearing.

Survey II has, conversely, essentially no incompleteness (i.e., $b = 1$), regardless of the evolutionary model. This implies that the channelization adopted for survey~II leads to an unbiased estimate of the true event rates.
\begin{figure}
	\includegraphics[width=\columnwidth]{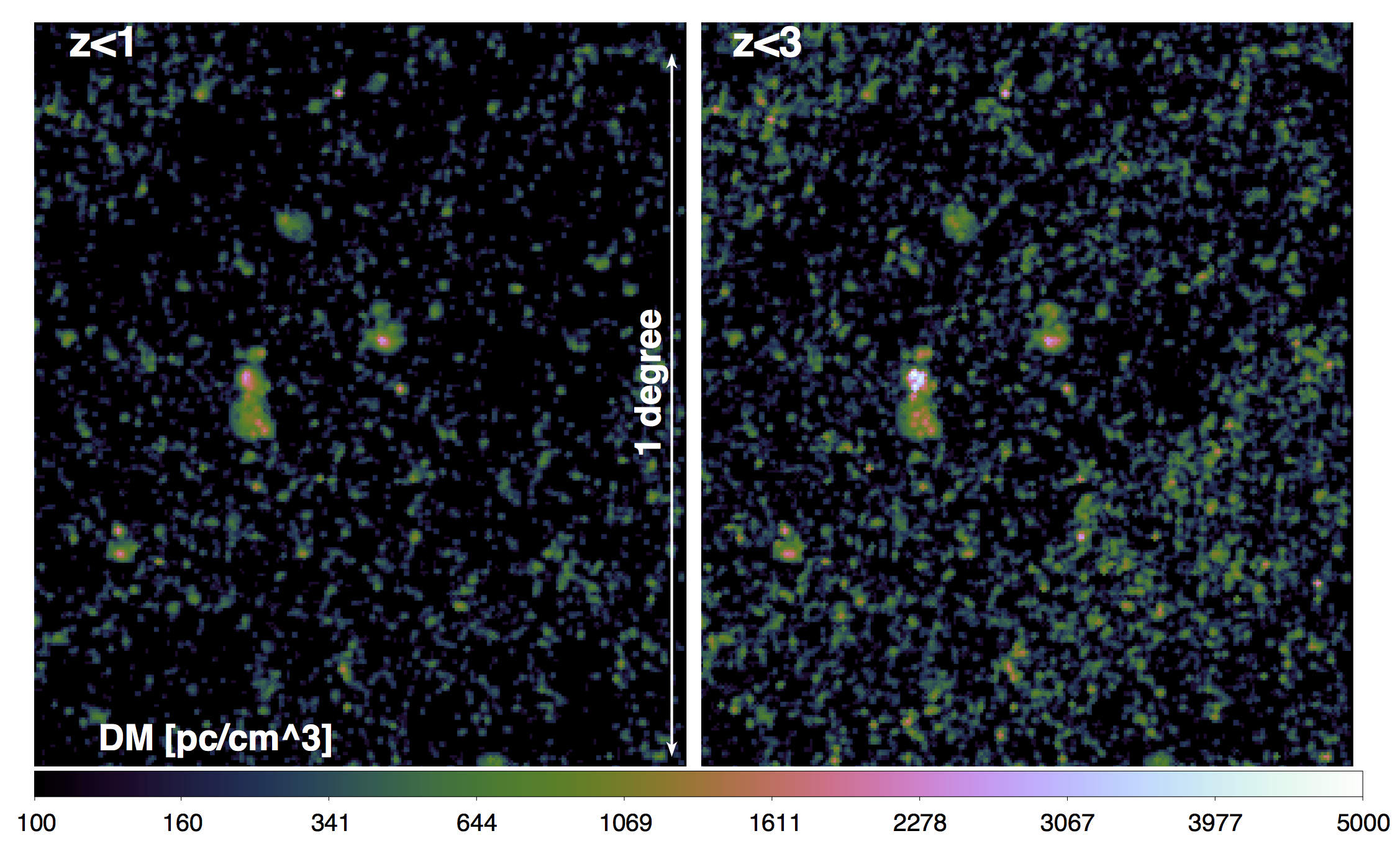}
	\includegraphics[width=\columnwidth]{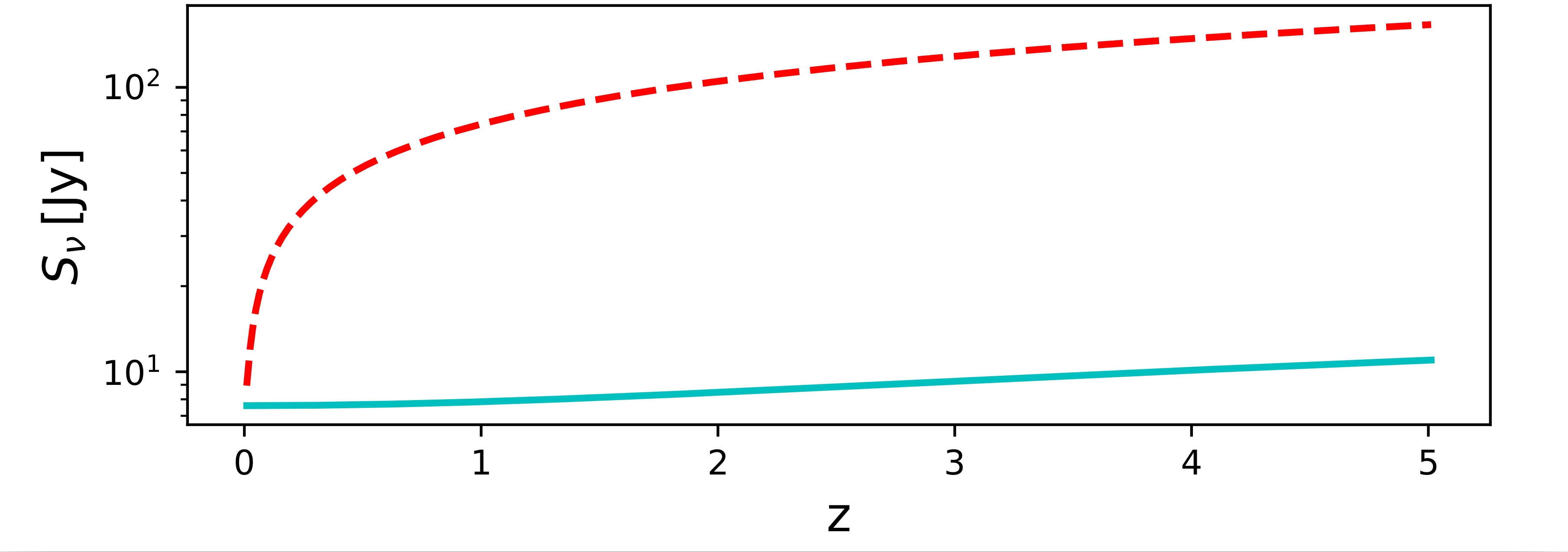}
    \caption{Upper panel: Simulated dispersion measure map from a cosmological simulation, for a full lightcone with $\approx 1^\circ$ aperture including the cosmic web up to $z=1$ (left) or $z=3$ (right).
    Lower panel: detection threshold as a function of the burst redshift, assuming a $z - {\rm DM}$ linear relation and a 1~ms burst duration, for survey~I (dashed-red line) and II (solid cyan line) respectively.}
    \label{fig:sim}
\end{figure}
The bias introduced by intra-channel smearing is redshift dependent as more distant events generally entail larger dispersion measures. Figure~\ref{fig:sim} displays two DM maps obtained from a cosmological simulation of the intergalactic medium \citep[see][for the simulation details]{2017CQGra..34w4001V} that clearly show that larger DM values corresponds to larger cosmological volumes and, therefore, higher redshift events. The lower panel of Figure~\ref{fig:sim} quantifies this effect using the linear DM-z relationship. For the survey I case, the dispersion of high redshift FRBs leads to a SNR decrease of about one order of magnitude at $z \sim 1$, that is, only the brightest events are observable at high redshift. For survey~II, conversely, the SNR only changes by $\sim 40\%$ up to $z =5$.   
We finally assessed how much survey~I and II constrain the FRB statistical properties.
We assumed that the probability density function ${\mathcal P}$ of observing $M$ events follows a Poissonian distribution \citep{vedantham17,amiri17}:
\begin{equation}
    {\mathcal P}(M \, | \, N(\alpha)) = C \, \frac{N(\alpha)^M e^{-N(\alpha)}}{M!},
\label{eq:poisson}
\end{equation}
where $N$ is the number of expected events and $C$ is a normalization factor, chosen  so that $ \int {\mathcal P}(\alpha) \, d\alpha = 1$. We assumed a power law shape for the event rates $N$: 
\begin{equation}
    N(\alpha) = 300 \, \left (\frac{S_\nu}{1 \, {\rm Jy}} \right)^{-\alpha} \times \rm FoV \times N_{\rm day} \,\,\,  \rm sky^{-1} day^{-1}, 
\label{eq:expected_events}
\end{equation}
where we used the event rate from \cite{CHIME2019b} as our pivotal value, that is, $300$ events brighter than 1~Jy observed in the $400-600$~MHz range.

The probability to find a slope smaller than $\alpha$ is thus given by the integral:
\begin{equation}
    P(<\alpha) = \int_{-\infty}^{\alpha} {\mathcal P}(M \, | \, N(\alpha')) \, d\alpha',
\end{equation}
while the probability of finding a slope greater than $\alpha$ is $1 - P(<\alpha)$.
\begin{figure}
	\includegraphics[width=\columnwidth]{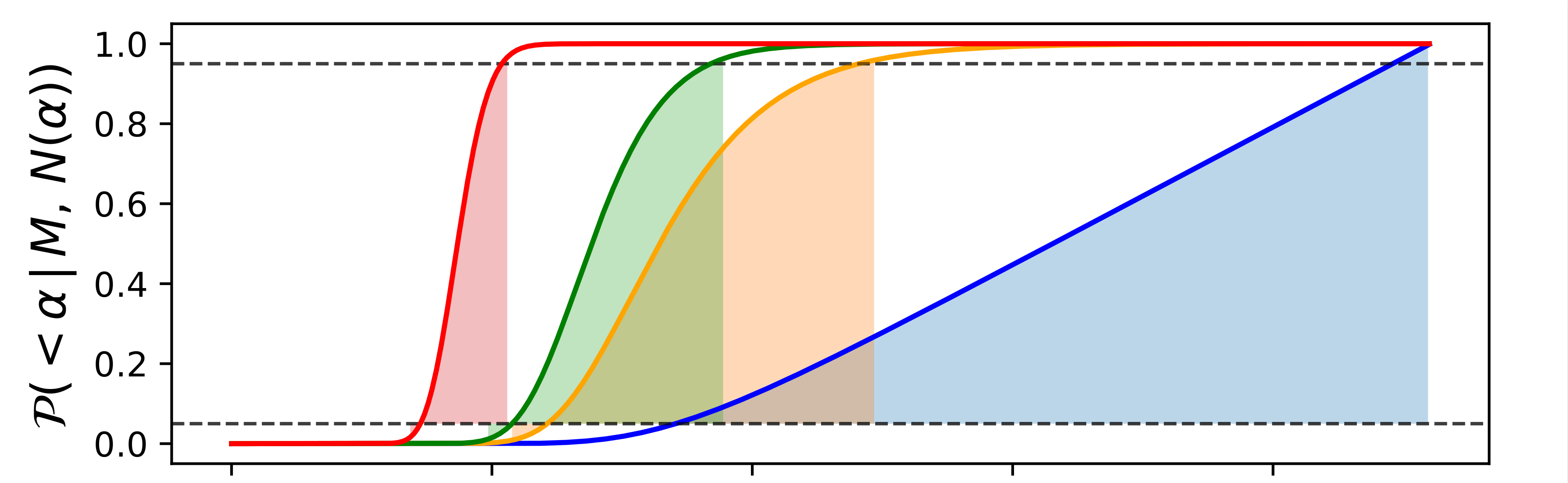}
	\includegraphics[width=\columnwidth]{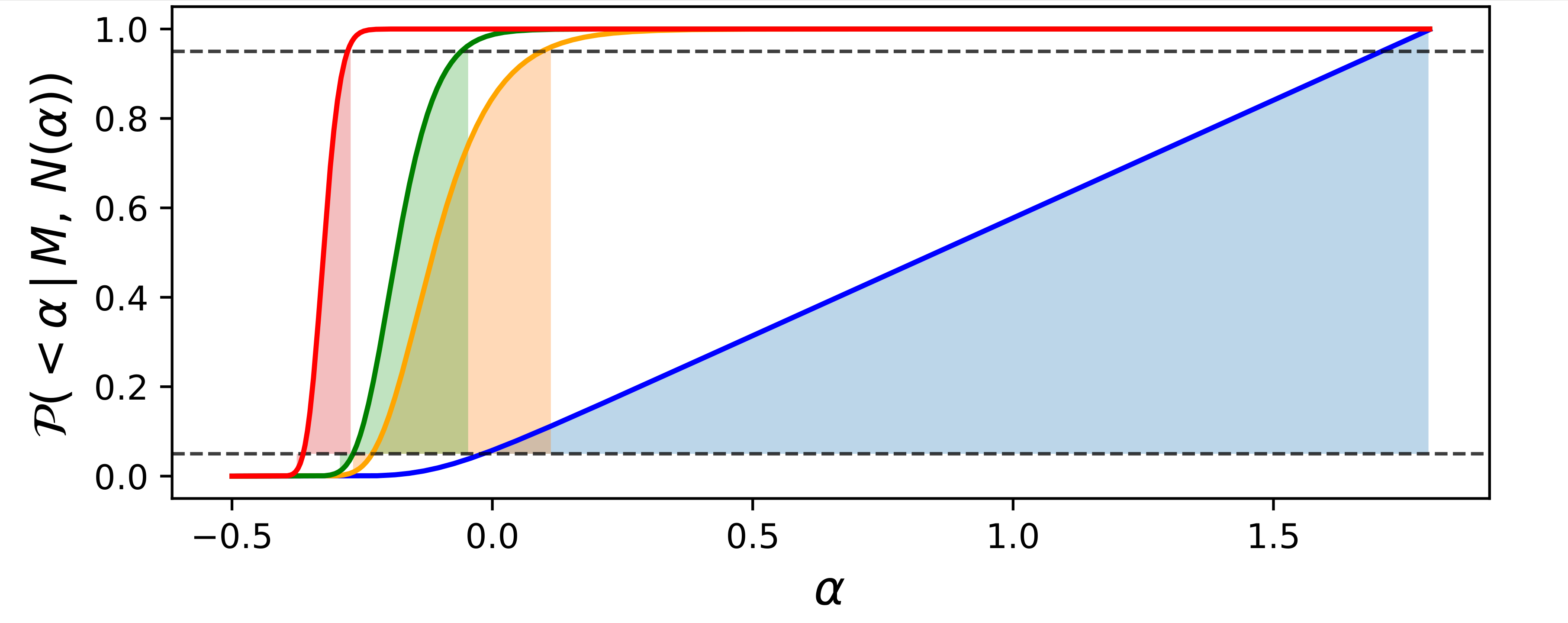}
    \caption{Constraints on the slope of the event rates for the survey~II (top ) and I (bottom panel) respectively. The probability is plotted as a function of 0 (blue), 1 (orange), 2 (green), 10 (red) observed events over $N (\alpha)$ expected events for 30 observing days (720~hours). 
    Horizontal dashed lines show the 5\% and 95\% confidence levels respectively.}
    \label{fig:constraints}
\end{figure}
Constraints on the slope of the event rates are shown in Figure~\ref{fig:constraints} for survey~I and II assuming a fiducial duration of 30~days.
Due to its larger FoV, survey~II will place better constraints on the event rate slope than survey~I. A non detection, in particular, will be able to rule out flat slopes, constraining $\alpha > 0.35$ at the 95\% confidence level. 

Assuming $\alpha = 1$ \citep[consistent with estimates at GHz frequencies;][]{vedantham17,amiri17}, we expect $\sim 40\,\rm sky^{-1}\, day^{-1}$ events above the survey detection threshold, leading to one detection every three days for survey~II. For survey~I we expect to detect one burst in $\sim  112 \,b^{-1}$ days, where the bias term incorporates the dependence upon the FRB evolutionary model due to DM smearing.

\section{Conclusions}\label{sec:conclusions}

In this paper we have described how the Northern Cross radio telescope is currently being equipped to carry out FRB surveys at 408~MHz.  
The current system uses a 16~MHz bandwidth divided in 21, 781~kHz wide channels and consists of eight cylinders whose inputs can be combined into either a single beam or four independent ones with a sub-ms time sampling. Tests of the digital and software back-end were carried out with six cylinders by observing the pulsar PSR\,B0329+54 from which the receiver ${\rm SEFD} = 1530$~Jy was derived. Based on the derived SEFD, we presented forecasts for FRB searches using two different models of their cosmological evolution for two cases, one which uses the current system with eight cylinders (survey~I) and an advanced one that uses sixty cylinders (survey~II), for which we assumed the back-end upgrades in terms of multi beam capabilities and finer channelization that are currently under development. For both cases the rms sensitivity is $\sigma_I = 760 \, (\tau/ {\rm ms})^{-0.5}$~mJy, with an instantaneous sky coverage of $9.4$~deg$^2$ and $350$~deg$^2$ respectively.

We found that the survey~I is expected to detect one FRB every $\sim 112$~days, although this rate suffers from smearing of high DM events and, therefore, depends upon the underlying FRB evolutionary model.
Survey~II is, conversely, immune from intra-channel smearing and is expected to yield one detection every three days, independently of the FRB model. Due to its large FoV, it is expected to probe FRBs up to $z \sim 5$ with an almost constant detection threshold.  
Based on the current low frequency event rates \citep{CHIME2019b}, survey~II will be able to constrain the slope $\alpha$ of the event rate. In particular, in the case of no detections, a 720~h campaign will yield $\alpha > 0.35$ at the 95\% confidence level. Assuming a fiducial slope $\alpha = 1$, we expect $\sim 40$~sky$^{-1}$~day$^{-1}$ events above a $10\sigma_I$ detection threshold, that is, one detection every three days. 

While the upgrade to carry out survey~II is ongoing, the current system is being used to monitor repeating FRBs 
and improved localization capabilities are being considered by deploying receiving systems at 408~MHz at the other Italian radio astronomical stations.

\section*{Acknowledgements}\label{acknowledgments}
We would like to thank an anonymous referee for useful comments. NTL and FV acknowledge financial support from the Horizon 2020 programme under the ERC Starting Grant "MAGCOW", no. 714196. NTL also thanks Lorenzo Gamba, Liam Connor and Manisha Caleb for helpful discussions. GB is in debt to Griffin Foster for his help in the early stages of the project.
AR gratefully acknowledges financial support by the research grant “iPeska” (P.I. Andrea Possenti) funded under the INAF national call Prin-SKA/CTA approved with the Presidential Decree 70/2016.
The ADU platform described in \S~\ref{sec:instrument} is the result of a collaboration between the Italian Institute for Astrophysics (INAF), University of Oxford, University of Malta, Science and Technology Facility Council (STFC, UK) and was supported by industrial partners.
The Northern Cross radio telescope is a facility of the University of Bologna operated under agreement by the Institute of Radio Astronomy of Bologna (INAF). 
The ENZO (enzo-project.org) simulation used for this work was produced on the Marconi-KNL Supercomputer at CINECA, under project no.INA17\_C4A28 with F.V. as PI.  We also acknowledge the usage of online storage tools kindly provided by the INAF Astronomical Archive (IA2) initiative (http://www.ia2.inaf.it). 
This research made use of Astropy,\footnote{http://www.astropy.org} a community-developed core Python package for Astronomy \citep{2013A&A...558A..33A}.




\bibliographystyle{mnras}
\bibliography{FRB} 





\bsp	
\label{lastpage}
\end{document}